\documentclass[journal]{IEEEtran}
\usepackage{amsfonts, amssymb}
\usepackage{latexsym}
\usepackage{graphicx}
\usepackage{algorithmic}
\usepackage{color}
\usepackage{setspace}
\usepackage[cmex10]{amsmath}
\usepackage{array}
\usepackage{eqparbox}
\usepackage{stfloats}
\usepackage{mdwmath}
\usepackage{mdwtab}
\usepackage{subfigure}
\usepackage{bm,stmaryrd}
\usepackage{bbm}
\usepackage{multirow}
\usepackage{multicol}
\usepackage{cite}
\newcommand{\indicator}[1]{\mathbbm{1}\left[ {#1} \right]}

\IEEEoverridecommandlockouts

\title{A Flexible Channel Coding Approach for Short-Length Codewords}

\author{Mikel~Hernaez,~\IEEEmembership{Student Member,~IEEE,}
 		Pedro~M.~Crespo,~\IEEEmembership{Senior Member,~IEEE,} \\and
        Javier~Del~Ser,~\IEEEmembership{Senior Member,~IEEE}}

\begin{document}
\maketitle
\begin{abstract}
This letter introduces a novel channel coding design framework for short-length codewords that permits balancing the tradeoff between the bit error rate floor and  waterfall region by modifying a single real-valued parameter. The proposed approach is based on combining convolutional coding with a $q$-ary linear combination and unequal energy allocation, the latter being controlled by the aforementioned parameter. EXIT charts are used to shed light on the convergence characteristics of the associated iterative decoder, which is described in terms of factor graphs. Simulation results show that the proposed scheme is able to adjust its end-to-end error rate performance efficiently and easily, on the contrary to previous approaches that require a full code redesign when the error rate requirements of the application change. Simulations also show that, at mid-range bit-error rates, there is a small performance penalty with respect to the previous approaches. However, the EXIT chart analysis and the simulation results suggest that for very low bit-error rates the proposed system will exhibit lower error floors than previous approaches.
\end{abstract}
%
%
\section{Introduction}\label{sec:intro}
Iteratively decodable (i.e. Turbo-like) channel codes such as Low Density Parity-Check (LDPC) \cite{LDPC} or Turbo codes \cite{Turbo}, have been widely shown to perform near capacity in point-to-point communications when used with codeword lengths beyond $10^6$. However, such codes can be impractical in many applications scenarios demanding low latencies (e.g. real-time video delivery), mainly due to their associated decoding complexity and limited technological resources of the underlying hardware. This rationale motivates the upsurge of research on short-length codes (i.e. codes with codewords of several hundreds to few thousands coded symbols) to the above scenarios.

However, when dealing with short-length codewords the capacity-approaching performance of LDPC or Turbo codes may severely degrade due to the high error floors obtained under this condition \cite{LDPC_EF,Turbo_EF}. Several contributions have focused on reducing these error floors such as \cite{Zheng10}, where a novel design technique is proposed to produce short-length parity-check matrices for LDPC codes that lead to low error floors. Indeed, as outlined in \cite{Zheng10} decreasing the error floor of LDPC codes is essential for potential applications such as data storage and deep-space communications, which require a Bit Error Rate (BER) as low as $10^{-15}$. However, there are some other applications where higher error floors can be allowed (e.g. a Quasi-Error-Free system in DVB-T broadcasting requires a maximum BER of $10^{-4}$). In these applications the code design paradigm is not to reduce the error floor to its minimum, but to achieve the imposed BER limit at a signal to noise ratio (SNR) as low as possible. In this context, the well-known design tradeoff between the error floor and the BER waterfall region existing in Turbo-like codes \cite{BICM-ID} becomes relevant. Unfortunately, balancing this tradeoff mainly depends on the design of the code itself (e.g. the parity-check matrix in LDPC codes), thus any change in the BER limit involves redesigning the code structure in its entirety.

In this paper we propose an alternative channel coding approach for short-length codewords that can be easily adapted to any BER needs of a given application without requiring a full redesign of the code. That is, it can easily switch between high error floors and low-SNR waterfall regions (as those imposed by e.g. DVB-T), or minimum error floors and higher-SNR waterfall regions (e.g. deep space communications). The proposed code hinges on adding extra parity bits, generated by linearly combining the encoded bits over a non-binary Galois field, to the already convolutionally-encoded bits. Moreover, unequal energy allocation is used between the convolutionally-encoded symbols and the linearly-combined symbols, and EXIT charts are used to analyze the performance of the associated decoder and to optimize the coefficients of the linear combination. Simulation results will show that the aforementioned tradeoff can be easily balanced by only varying a single parameter, namely, the unequal energy allocation parameter $\Lambda$, hence the switch between regions can be done without changing the encoder and decoder.

The manuscript is organized as follows: Section \ref{sec:code} presents the system model, the proposed encoder and the corresponding decoder, whereas an analysis of the code based on EXIT charts is performed in Section \ref{sec:EXIT}. Next, Section \ref{sec:res} discusses the obtained Monte Carlo simulation results and ends the paper by drawing some concluding remarks.
%
%
\section{Proposed Code}\label{sec:code}

In order to ease the understanding of the proposed code some definitions are first introduced. Consider the set of all $2^q$ polynomials $\rho(z)$ of degree $q-1$ with coefficients lying in $GF(2)$ (the binary Galois field). Let $g(z)$ be a prime polynomial (i.e., monic and irreducible polynomial) of order $q$. Then, this set becomes a finite field, $GF(2^q)$, by defining the addition $\oplus$ and multiplication $\otimes$ rules as the $\hspace{-3mm}\mod{g(z)}$ remainder of the sum and product of two polynomials, respectively. Notice that, since the $\hspace{-3mm}\mod{g(z)}$ addition rule is just a componentwise addition of coefficients in $GF(2)$, $GF(2^q)$ under addition is isomorphic to the vector space $(GF(2))^q$ of binary $q$-tuples with $\hspace{-3mm}\mod{2}$ elementwise addition denoted hereafter as $\wedge$. Therefore, there is a one-to-one mapping $\psi_q: (GF(2))^q \rightarrow GF(2^q)$ defined as $\psi_q(a_0,\ldots, a_{q-1})= \sum_{k=0}^{q-1} a_{k}z^k$ such that $\psi_q(\mathbf{a}) \oplus \psi_q( \mathbf{b})= \psi_q (\mathbf{a} \wedge \mathbf{b})$, where $\mathbf{a},\mathbf{b}\in (GF(2))^q$. In addition, we index the elements $\rho_i\in GF(2^q)$, $i\in\{0,\ldots,2^q-1\}$ by the base-10 notation of the corresponding binary tuple $(a_0,\ldots, a_{q-1})$. In the following we refer as \emph{non-binary} symbols to the elements of the finite field $GF(2^q)$.
\begin{figure}[!h]
\centering
  \includegraphics[width=\columnwidth]{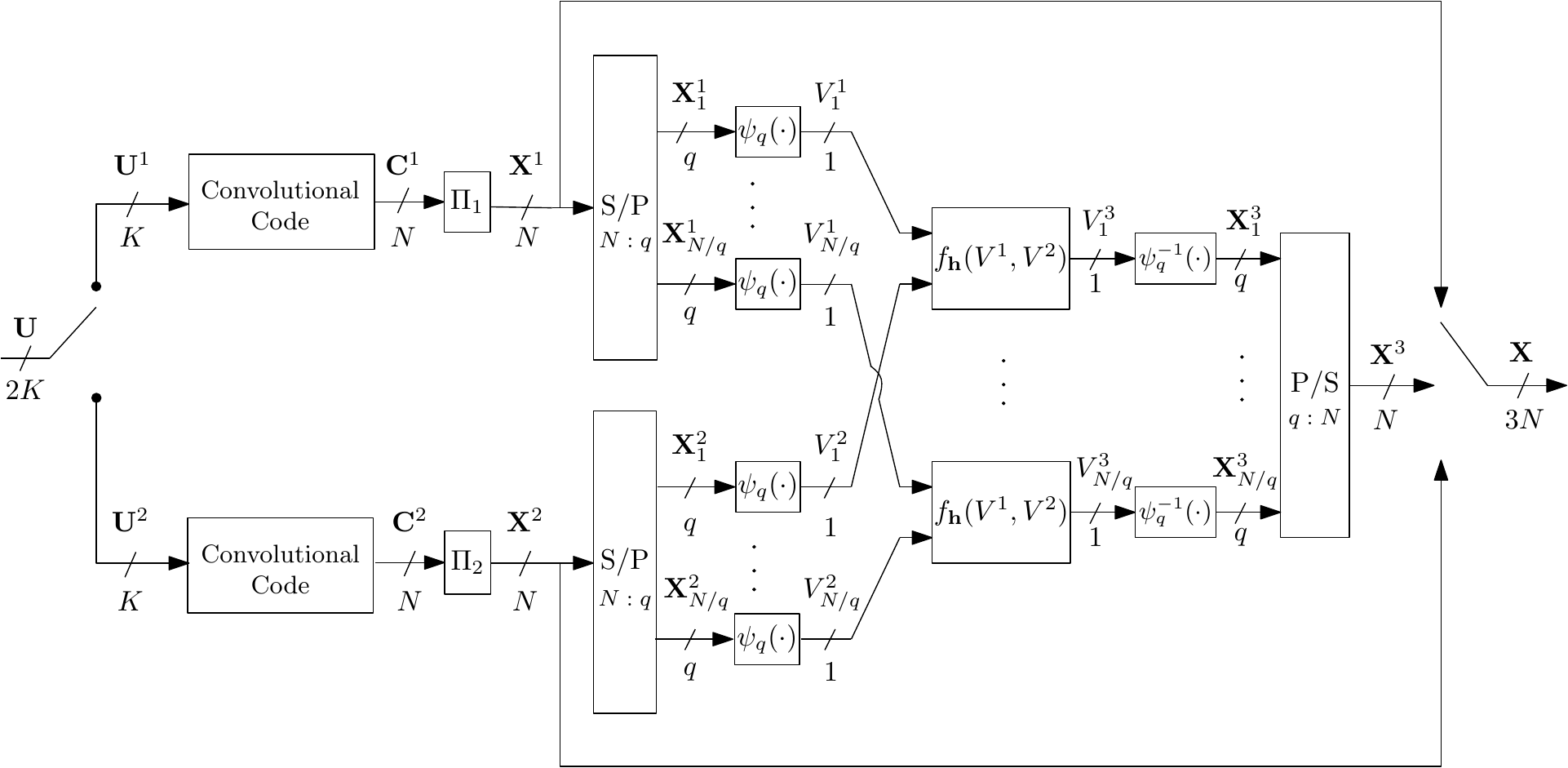}
  \caption{Encoder associated to the proposed code.}\label{fig:encoder}
\end{figure}

We consider a point-to-point scenario consisting of a binary unit-entropy information source $\mathcal{S}$, which generates blocks $\mathbf{U}\in \{0,1\}^{2K}$. As depicted in Figure \ref{fig:encoder}, the sequence $\mathbf{U}$ is divided into two sequences $\mathbf{U}^1$ and $\mathbf{U}^2$ of length $K$, which are channel-coded by a terminated convolutional code, producing the codewords $\mathbf{C}^m\triangleq \{C^m_t\}_{t=1}^N \in \{0,1\}^N$, with $m\in\{1,2\}$. Then, each codeword is interleaved yielding the interleaved codeword $\mathbf{X}^m=\Pi_m\left(\mathbf{C}^m\right)$, where $\Pi_1$ and $\Pi_2$ are two different spread interleavers with a spread factor equal to $q$. Next, each of the interleaved coded sequences $\mathbf{X}^1$ and $\mathbf{X}^2$ is split into $q$-length sub-sequences $\{\mathbf{X}^1_l\}_{l=1}^{N/q}\triangleq\{X^1_{l,1},\ldots,X^1_{l,q}\}_{l=1}^{N/q}$ and $\{\mathbf{X}^2_l\}_{l=1}^{N/q}$, respectively. We denote as $V_l^m=\psi_q(\mathbf{X}^m_l)\in GF(2^q)$ the non-binary symbol associated to the corresponding sub-sequence: the non-binary symbol $V_l^3$ is computed as the linear combination of the non-binary symbols $V_l^1$ and $V_l^2$, i.e.
\begin{equation}\label{eq:lc}
\hspace{-.5mm}V_l^3\hspace{-.5mm}\triangleq\hspace{-.5mm}\psi_q
\hspace{-.25mm}\left(\psi_q^{\mbox{-}1}\hspace{-.5mm}\left(h^1
\hspace{-.25mm} \varotimes V_l^1\right)\hspace{-.5mm}\wedge \psi_q^{\mbox{-}1}\hspace{-.5mm}\left(h^2\hspace{-.25mm} \varotimes  V_l^2\right)\right)\hspace{-.5mm}\triangleq\hspace{-.5mm} f_\mathbf{h}(V_l^1,V_l^2),
\end{equation}
where $\mathbf{h}=(h^1,h^2)$, $h^m\in \{\rho_i\}_{i=1}^{2^q-1}$ represents the coefficients used in the linear combination. Each sub-sequence $\mathbf{X}^3_l$ is computed from the associated non-binary symbols $V_l^3$ as $\mathbf{X}^3_l=\psi^{-1}_q(V_l^3)$. In the following, we refer as Linear Combination (LC) code to the rate-2/3 code formed by the sub-codewords $(\mathbf{X}_l^1, \mathbf{X}_l^2, \mathbf{X}_l^3)$. Thus, there are $N/q$ parallel LC codes. Finally, the sequence $\mathbf{X}^3$ is given by $\mathbf{X}^3=\{\mathbf{X}^3_l\}_{l=1}^{N/q}$, and the final codeword $\mathbf{X}=\{X_t\}_{t=1}^{3N}$ is formed by $\mathbf{X}=(\mathbf{X}^1, \mathbf{X}^2, \mathbf{X}^3)$. Thus, the overall code rate is given by $2K/3N$.

We apply unequal energy allocation between the sub-codewords associated to the terminated convolutional codes ($\mathbf{X}^1$ and $\mathbf{X}^2$) and the sub-codeword associated to the linear combination ($\mathbf{X}^3$). We denote as\footnote{$\mathrm{E}[\cdot]$ stands for expected value.} $E_s^{cc}=\mathrm{E}[(X^m)^2]$ the energy of the sub-codewords $\mathbf{X}^1$ and $\mathbf{X}^2$, whereas $E_s^{lc}=\mathrm{E}[(X^3)^2]$ and $\overline{E}_s=\mathrm{E}[X^2]$ denote the energy of $\mathbf{X}^3$ and $\mathbf{X}$, respectively. Furthermore, without loss of generality, we assume $E_s^{lc}=\lambda E_s^{cc}$, with $\lambda>0$ and define $\Lambda=10\log{\lambda}$ (dB). Thus, we have
\begin{equation}
\begin{array}{cc}
E_s^{cc}=\frac{3}{2+\lambda}\overline{E}_s, & E_s^{lc}=\frac{3\lambda}{2+\lambda}\overline{E}_s.\\
\end{array}
\end{equation}

This results in the modulated symbols $S_t=2X_t-1$, with $t=1,\ldots, 3N$, and average energy per symbol $\overline{E}_s$. The received symbol per real dimension at the receiver is given by $Y_t=S_t+N_t$, where $\{N_t\}_{t=1}^{3N}$ are modelled as real Gaussian i.i.d. random variables with zero mean and variance $N_0/2$.
%
%
\subsection{Proposed Decoder} \label{sec:Decoder}
The destination receives the channel outputs $\mathbf{Y}=(\mathbf{Y}^1,\mathbf{Y}^2,\mathbf{Y}^3)$, where the subsequence $\mathbf{Y}^j$ (with $j\in\{1,2,3\}$) represents the channel outputs associated to encoded symbols $\mathbf{X}^j$. The aim of the decoder is to estimate the source binary symbols $\mathbf{\widehat{U}}=(\{\widehat{U}_k^1\}_{k=1}^K,\{\widehat{U}_k^2\}_{k=1}^K)$ so that the conditional probability $P(u^m_k|\mathbf{y})$  ($m\in\{1,2\}$) -- which is obtained by marginalizing the joint conditional probability $P(\mathbf{u}^m|\mathbf{y})$ -- is maximized. This marginalization is efficiently computed by applying the Sum-Product Algorithm (SPA, see \cite{SPA}) to the factor graph describing $P(\mathbf{u}^m|\mathbf{y})$, which is shown in Figure \ref{fig:decoder}. Observe that such an overall factor graph is composed by three sub-factor graphs: two describing the convolutional codes, and a third one describing the LC code. Since this factor graph has loops, the SPA is iteratively run between the sub-factor graphs corresponding to the LC code and the convolutional codes. After a fixed number of iterations $\mathcal{I}$, the probability $P(u^m_k|\mathbf{y})$ based on which $\widehat{U}_k^m$ is computed results proportional to
\begin{multline}
\sum_{\sim u^m_k} \hspace{-0.7mm}T_k(\texttt{s}_k^m,u^m_k,\mathbf{c}_k^m,\texttt{s}_{k+1}^m)\alpha(\texttt{s}_k^m)\beta(\texttt{s}_{k+1}^m)
\hspace{-2.5mm}\prod_{t:\ c_t^m\in \mathbf{c}^m_k} \hspace{-2.5mm} \gamma(c_t^m), \nonumber
 \end{multline}
where $\sim u_k^m\triangleq \{u_{k'}^m\}_{\forall k'\neq k}$; $\alpha$ and $\beta$ are the messages passed from the adjacent state nodes to the factor node $T_k$ given by the transitions in the Trellis describing the convolutional code; $\mathbf{c}_k^m=\{c_t^m\}_{t=(k-1)N/K}^{kN/K}$ (with $k\in\{1,\ldots,K\}$) represents the coded bits associated to $u_k^m$; and $\gamma(\cdot)$ are the likelihoods passed from the variable nodes $c_t^m$ to $T_k$. These likelihoods depend on the messages passed by the LC-check nodes associated to the interleaved binary symbol $x_{\Pi^{-1}_m(t)}^m$, i.e. $\gamma(c^m_t)=\gamma(x^m_{\Pi^{-1}_m(t)})$, where $\gamma(x^m_t)\propto p(\mathbf{y}_l\vert x_t^m)$ and $\mathbf{y}_l\triangleq(\mathbf{y}_l^{1}, \mathbf{y}_l^{2}, \mathbf{y}_l^{3})$ (i.e. those components of $\mathbf{Y}$ associated to the LC check node $LC_l$). It can be shown that such likelihoods can be further factorized as
\begin{multline}\label{eq:PYX}
p(\mathbf{y}_l\vert x^m_{l,i})=\sum_{\sim x_{l,i}^m,v_l^m}\indicator{v_l^m=\psi_q(x^m_{l,1},\ldots,x^m_{l,q})}\cdot\\ p(\mathbf{y}_l^m\vert v_l^m)\prod_{i'\neq i} P^a(x^m_{l,i'})P^{\mbox{\tiny LC}}(v_l^m),
\end{multline}
where $i\in\{1,\ldots,q\}$, $\indicator{\cdot}$ is an indicator function taking value $1$ if its argument is true and $0$ otherwise, and
\begin{multline}\label{eq:P_MARC}
P^{\mbox{\tiny LC}}(v_l^m)\triangleq\sum_{v_l^{\overline{m}},v_l^3}\indicator{v_l^3=f_\mathbf{h}(v_l^1,v_l^2)} p(\mathbf{y}_l^{\overline{m}}\vert v_l^{\overline{m}}) \\
\cdot P^a(v_l^{\overline{m}})p(\mathbf{y}_l^3\vert v_l^3),
\end{multline}
with $\overline{m}=3-m$. Since the non-binary symbols can be also expressed by the modulated symbols $\{S^j_{l,i}\}_{i=1}^q\in\{\pm 1\}^q$ as $V_l^j= \psi_q\left(\left\{(1+S_{l,i}^j)/2\right\}_{i=1}^q\right)\triangleq\theta_q
\left(\{S_{l,i}^j\}_{i=1}^q \right)$, we have that
\begin{multline}\label{eq:PYjV}
p(\mathbf{y}_l^j\vert v_l^j)\propto\sum_{s^j_{l,1},\ldots,s^j_{l,q}}\indicator{v_l^j=\theta_q(s^j_{l,1},\ldots,s^j_{l,q})}\\
\cdot \prod_{i=1}^q \exp\left(\frac{\mbox{-}\left|y_{l,i}^j-s_{l,i}^j\right|^2}{N_0}\right).
\end{multline}

In light of the above factorization, it is clear to see that the factor graph of the LC code is in turn composed of $N/q$ parallel and identical sub-factor graphs $LC_l$, which are depicted, for $l=1$ and $l=N/q$, in Figure \ref{fig:decoder}.
\begin{figure}[h!]
\centering
\includegraphics[width=0.9\columnwidth] {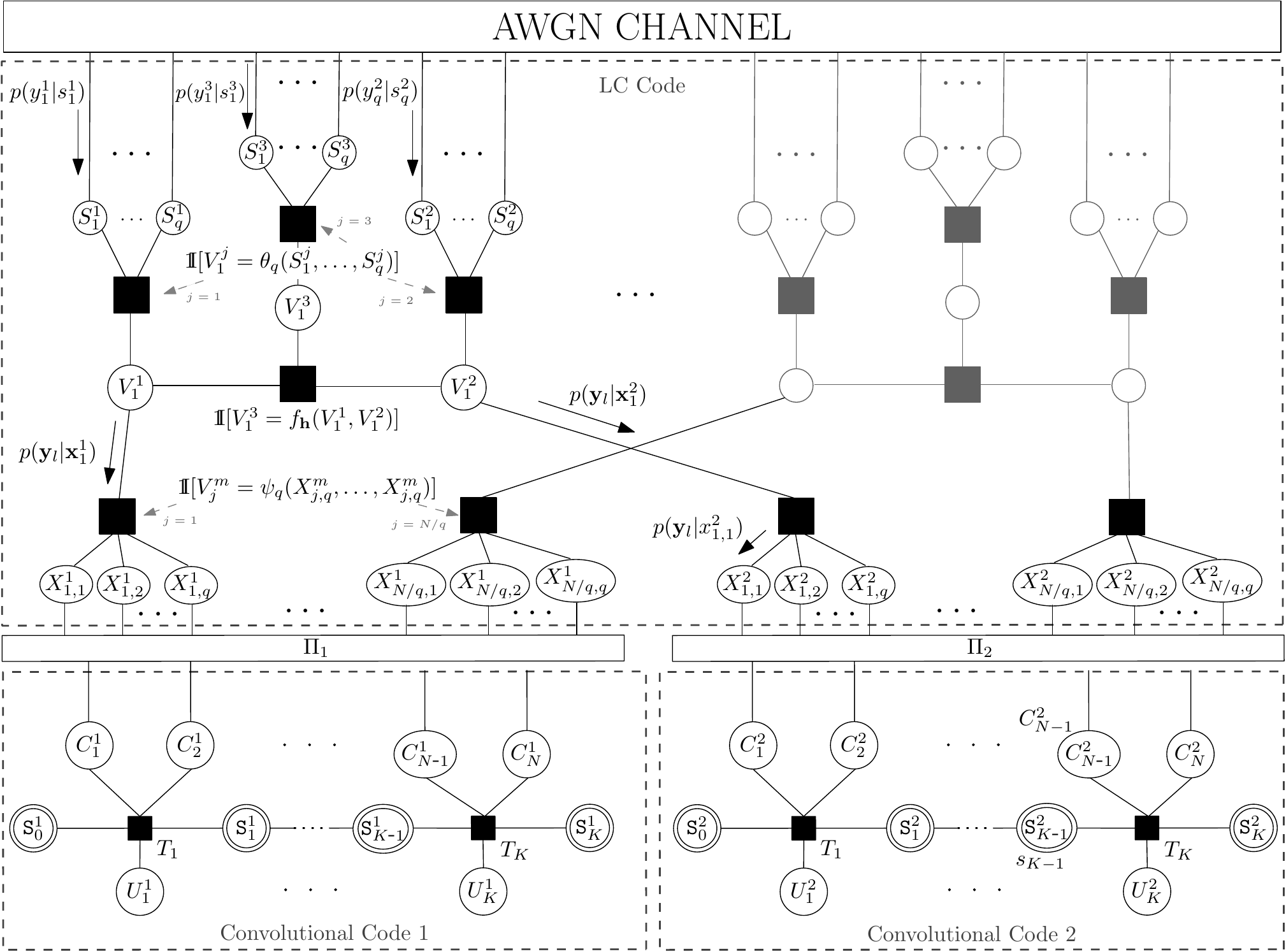}\label{fig:decoder}
    \caption{Factor graph of the proposed decoder.}\label{fig:decoder}
\end{figure}
%
%
\section{Code Analysis through EXIT Charts} \label{sec:EXIT}
The EXtrinsic Information Transfer (EXIT)  \cite{EXIT} function of a code is defined as the relationship between the mutual information at the input of the decoder (commonly denoted as $I_a$) and the extrinsic mutual information $I_e$ at its output, i.e. $I_e=T(I_a)$. For an iterative code, the chart plotting the transfer functions of the compounding sub-codes is called EXIT chart, and is known to be a powerful tool for designing and predicting the behavior of iterative codes (see \cite{EXIT} and reference therein).

We denote the transfer function of the LC-code for a given $q$ and $\mathbf{h}$ as $I_e^{\mbox{\tiny NC} }=T_{\mathbf{h}}^q(I_a^{\mbox{\tiny NC} })$. Notice, that for $I_a^{\mbox{\tiny NC} }=0$ the value of $I_e^{\mbox{\tiny NC} }=T_{\mathbf{h}}^q(0)$  will also depend on the Signal-to-Noise Ratio (SNR), since the channel observations $\mathbf{y}$ are used by the LC-code (last term of \eqref{eq:PYjV} and Fig. \ref{fig:decoder}). As the mutual information at the input of the channel decoders $I_a^{\mbox{\tiny CC}}$ is equal to $I_e^{\mbox{\tiny NC} }$, the extrinsic mutual information at the output of the convolutional decoders is given by $I_e^{\mbox{\tiny CC} }=T^{\mbox{\tiny CC} }(I_e^{\mbox{\tiny NC} })$. Thus, for a successful decoding procedure, there must be an open gap between both EXIT curves so that the iterative decoding can proceed from $I_e^{\mbox{\tiny CC} }=0$ to $I_e^{\mbox{\tiny CC} }= 1$. When both transfer functions cross, the iterative process will stop at a given extrinsic mutual information of the source bits $I_e^{\mbox{\tiny CC} }< 1$.

\begin{figure}[h!]
\centering
  \subfigure[]{ \includegraphics[width=0.85\columnwidth]{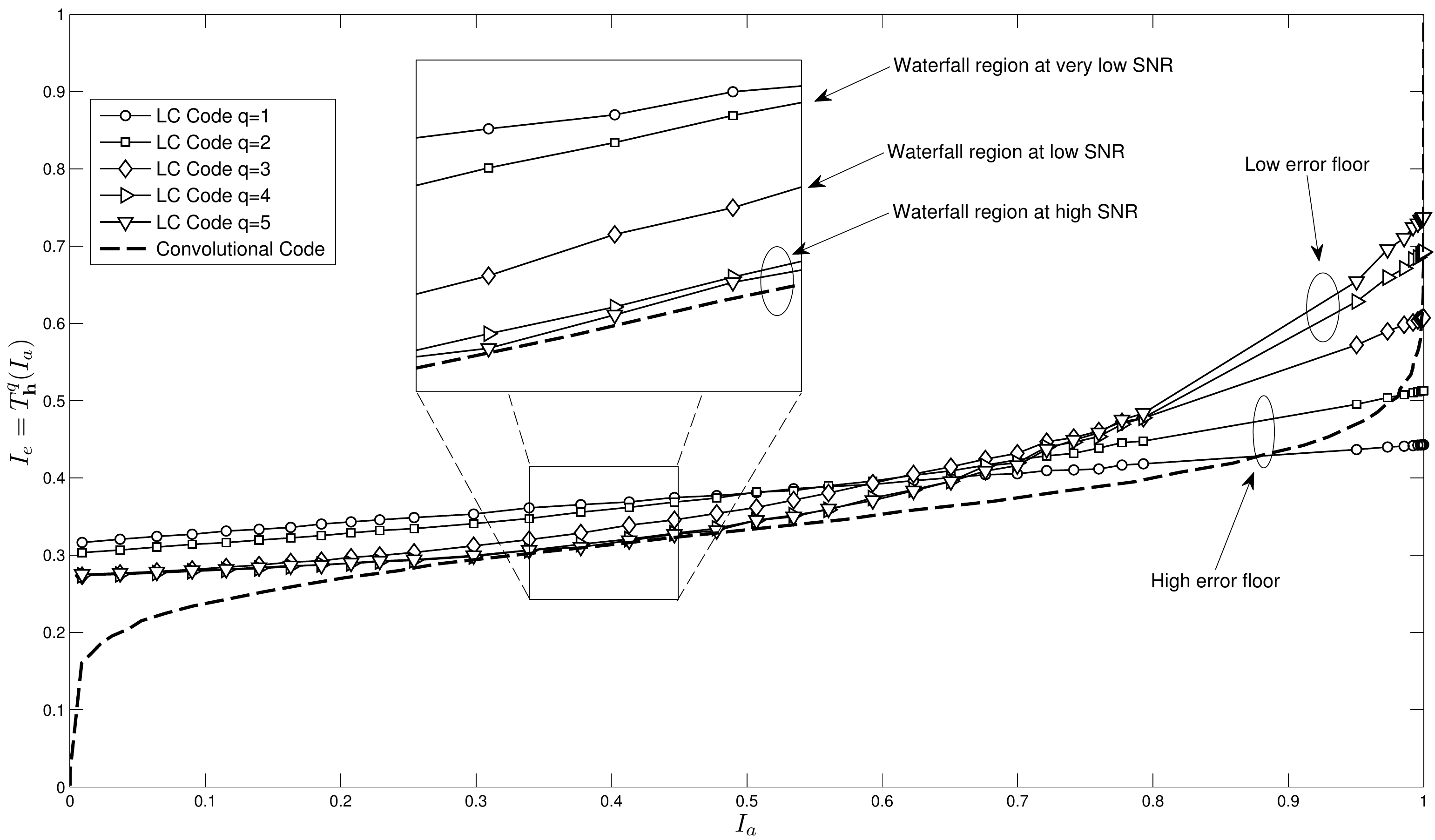}\label{fig:EXITq}}\\
  \subfigure[]{ \includegraphics[width=0.85\columnwidth]{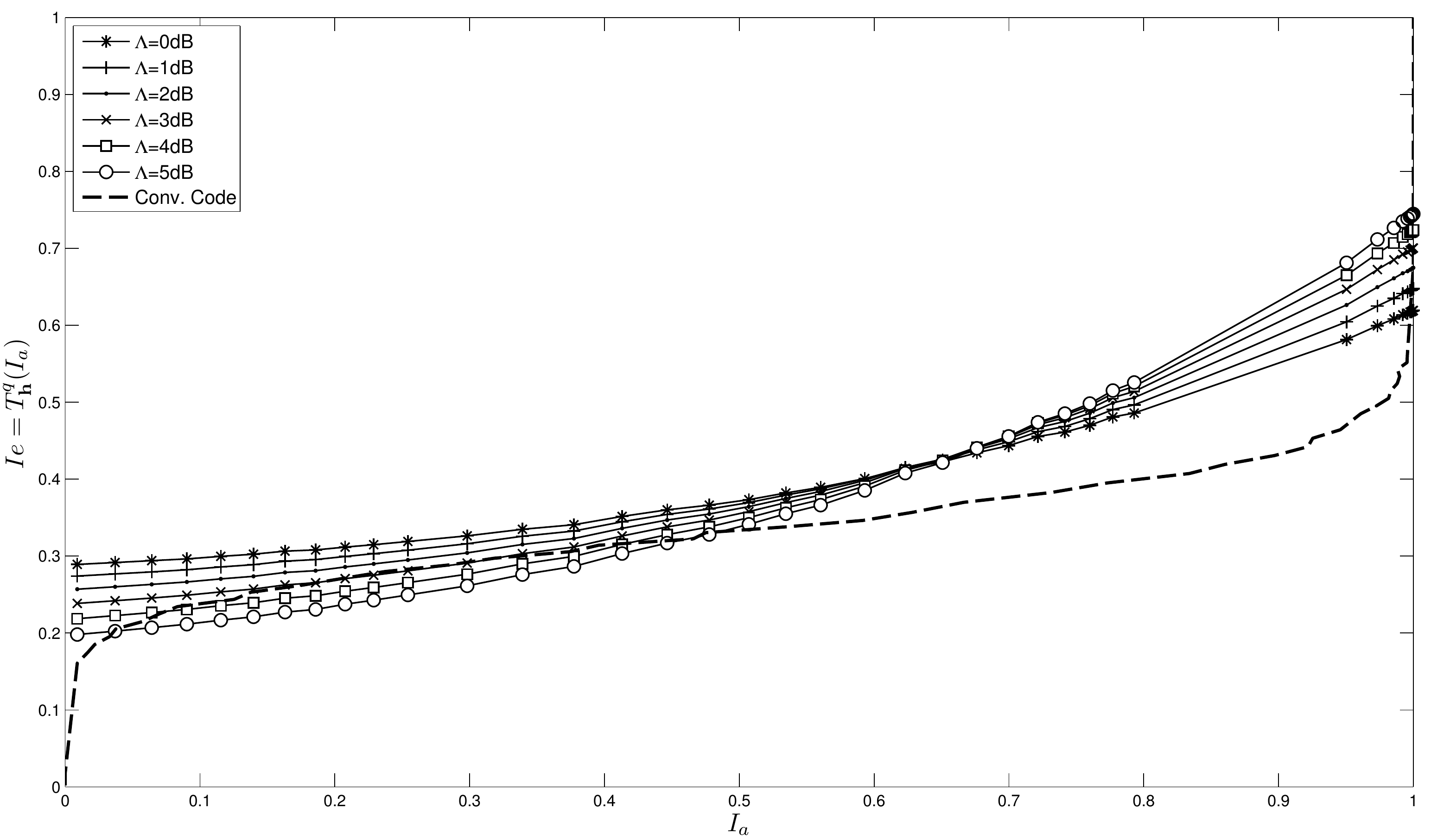}\label{fig:EXITL}}\\
  \caption{EXIT Chart of different LC codes and a rate-1/2 Convolutional Code for different values of $q$ (a) and $\Lambda$ (b).}\label{fig:EXIT}
\end{figure}

Since the transfer functions are monotonically increasing functions, the higher the value of $T_{\mathbf{h}}^q(0)$ is, the smaller the required SNR  to open a gap will be. On the other hand, if no crossing between curves has been produced, the higher the value of $T_{\mathbf{h}}^q(1)$ is, the closer $I_e^{\mbox{\tiny CC} }$ will be to 1 yielding lower error floors (see Fig. \ref{fig:EXIT}). Therefore, we are interested on having high values of both $T_{\mathbf{h}}^q(1)$ and $T_{\mathbf{h}}^q(0)$.

However, due to the Area Theorem of EXIT charts \cite{EXIT}, which states that the area under the transfer function depends only on the rate of the encoder, high values of $T_{\mathbf{h}}^q(1)$ yield to low values of $T_{\mathbf{h}}^q(0)$ and vice versa. Therefore, for a given $q$ we select a $\mathbf{h}$ that maximizes $T_{\mathbf{h}}^q(1)\cdot T_{\mathbf{h}}^q(0)$. For the sake of simplicity, this is done through exhaustive search over the $(q^2-1)^2$ possible values of $\mathbf{h}$.

Next, we justify the reason why the different linear combinations have a greater impact on the value of $T_{\mathbf{h}}^q(1)$ than on $T_{\mathbf{h}}^q(0)$. When $I_a^{\mbox{\tiny NC} }=0$ all the information at the LC-Code comes equally from the channel observations associated to $\mathbf{X}^1$, $\mathbf{X}^2$ and $\mathbf{X}^3$. On the other hand, when $I_a^{\mbox{\tiny NC} }=1$, only the information provided by the channel observations associated to $\mathbf{X}^3$ is relevant, as the information regarding $\mathbf{X}^1$ and $\mathbf{X}^2$ is fully supplied by the convolutional decoders.

Furthermore, variations on the unequal energy allocation parameter $\Lambda$ are expected to affect to $T_{\mathbf{h}}^q(1)$ and $T_{\mathbf{h}}^q(0)$ similarly, given that $\Lambda$ operates identically on $\mathbf{X}^1,\ \mathbf{X}^2$ and $\mathbf{X}^3$.

Figure \ref{fig:EXITq} shows that as $q$ increases, the value of $T_{\mathbf{h}}^q(1)$ increases yielding lower error floors, but  obtaining BER waterfall regions at higher SNRs. However, as shown in Fig. \ref{fig:EXITL}, the tradeoff between the SNR at which waterfall regions occur and its associated BER floor can be balanced through the selection of $\Lambda$.  By increasing $\Lambda$ the value of $T_{\mathbf{h}}^q(1)$ is increased, leading to low error floors. However, the value of $T_{\mathbf{h}}^q(0)$ is decreased, hence obtaining BER waterfall regions at higher SNRs.
%
%
\section{Simulation Results}\label{sec:res}
In order to assess the performance of the proposed code, several Monte Carlo simulations have been run using 6-memory-block $[554, 774]_8$ non-systematic rate-1/2 convolutional codes\footnote{The subindex $8$ in the definition of the code stands for \emph{octal}.}. A zero-bit tail is appended at the source sequence in order to terminate the convolutional codes properly. Note that this added bit tail reduces the overall spectral efficiency of the system, specially with very short-length codewords. Therefore, it must be taken into account when computing the gap to the Shannon Limit. The interleavers $\Pi_m\left(\cdot\right)$ have been randomly generated and are independent from each other. Finally, we have considered binary PAM modulation with transmitted blocks of $3N=2000+q$ real symbols, and $\mathcal{I}=20$ iterations of the decoding algorithm. The decoding is stopped at every simulated SNR point when $100$ errors have been obtained.

\begin{figure}
\centering
  \subfigure[$q=2$.]{ \includegraphics[width=0.48\columnwidth]{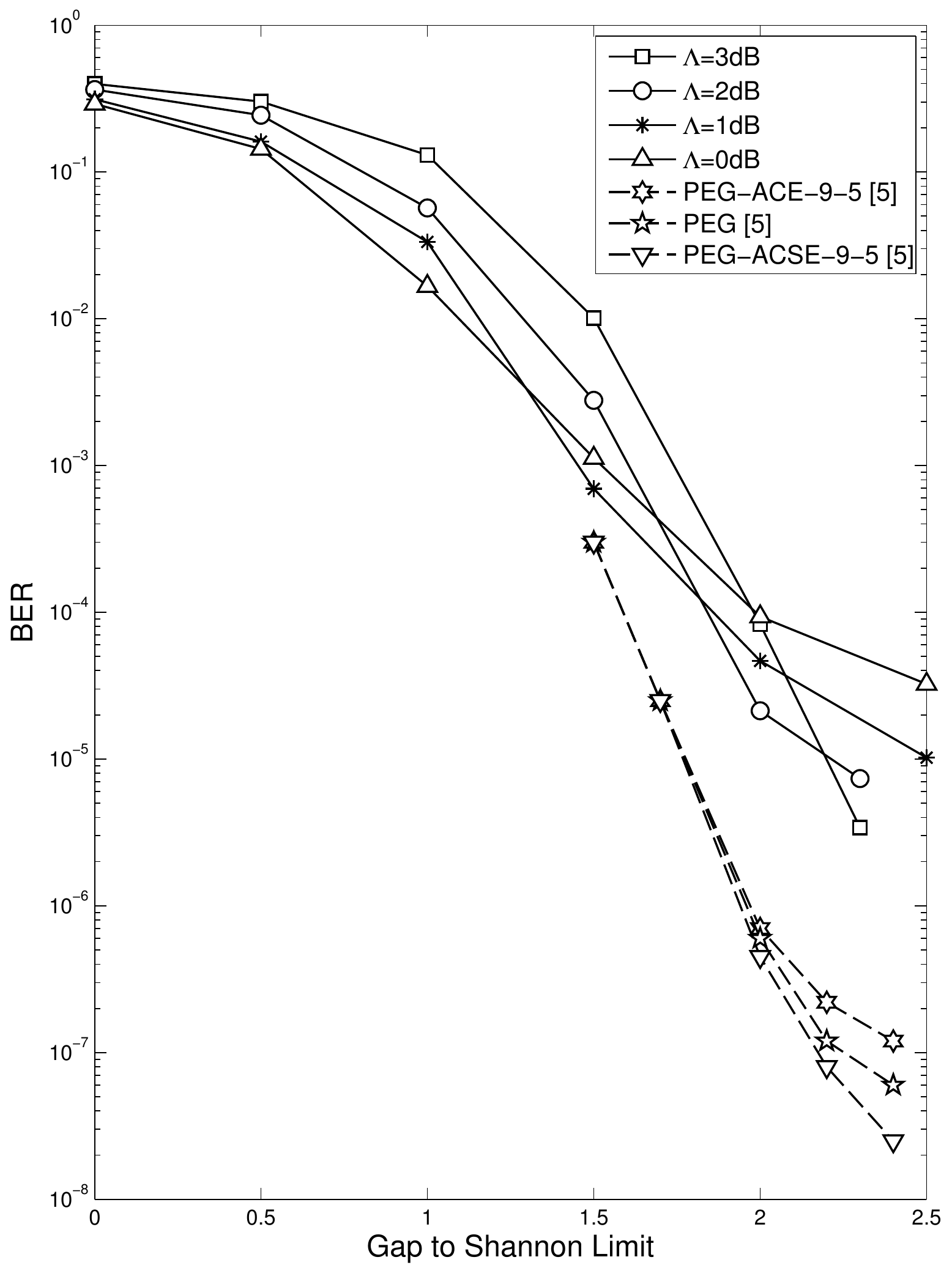}\label{fig:BERq2}}
  \subfigure[$q=3$.]{ \includegraphics[width=0.48\columnwidth]{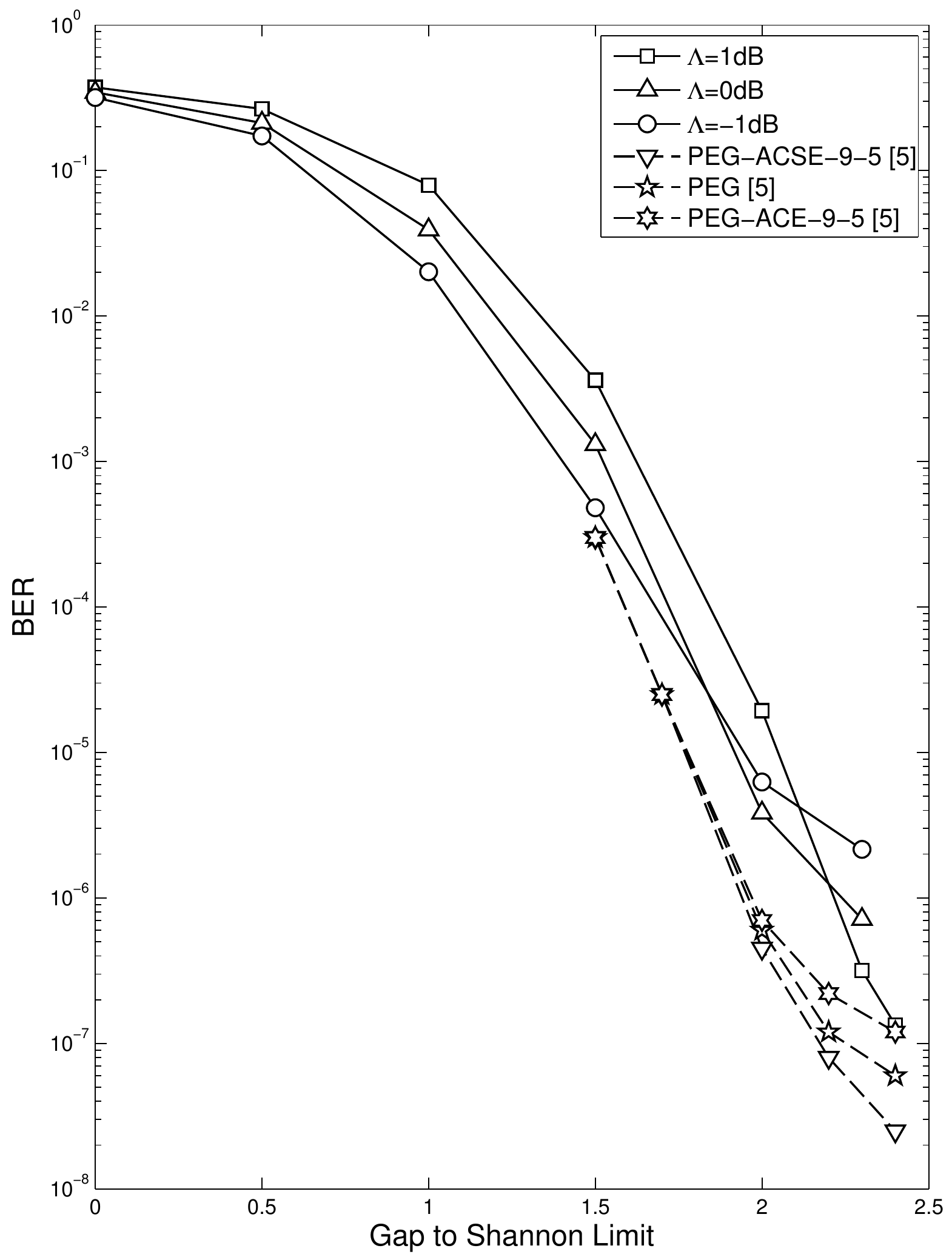}\label{fig:BERq3}}\\
  \subfigure[$q=4$.]{ \includegraphics[width=0.48\columnwidth]{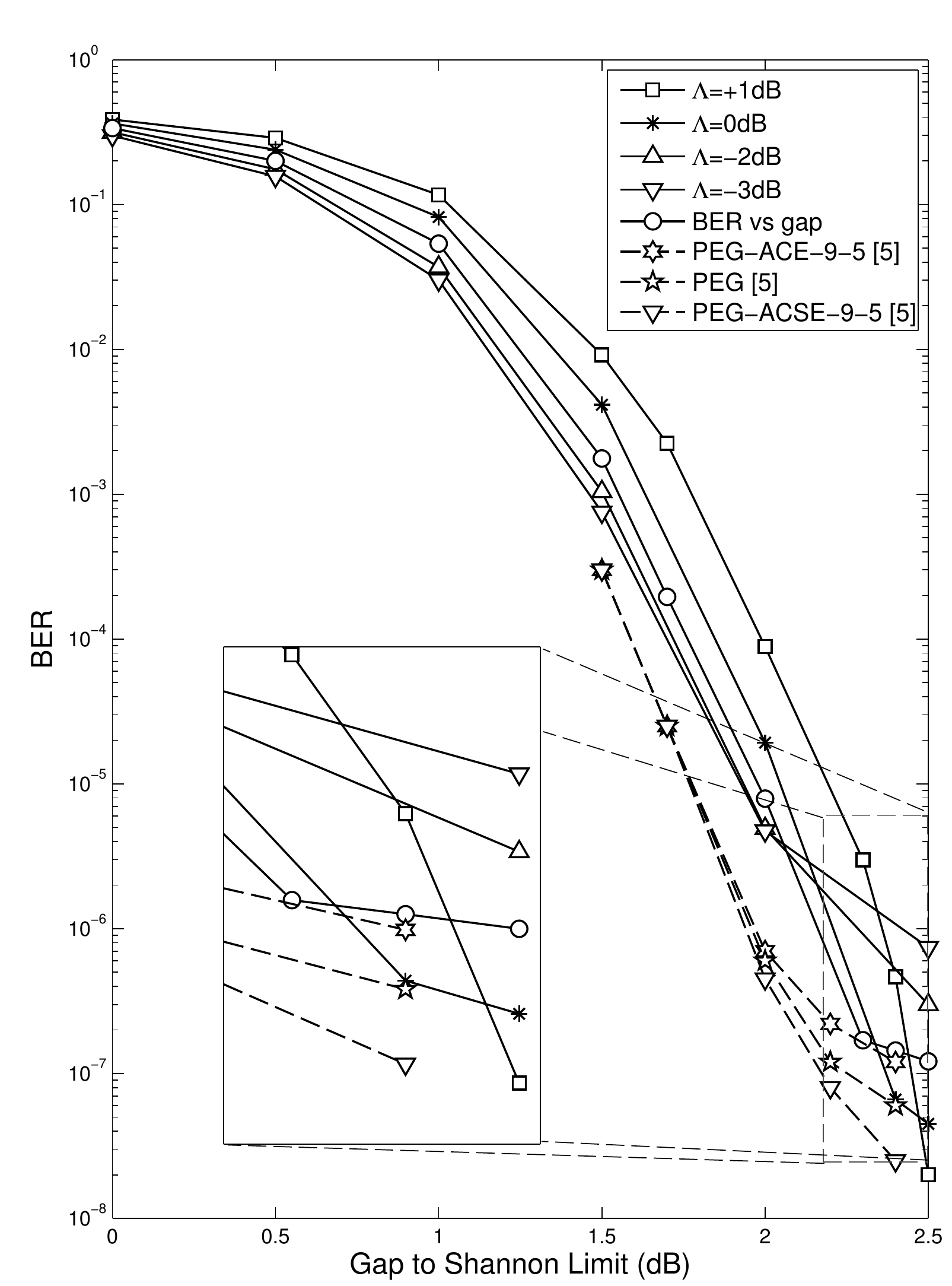}\label{fig:BERq4}}
  \subfigure[$q=5$.]{ \includegraphics[width=0.48\columnwidth]{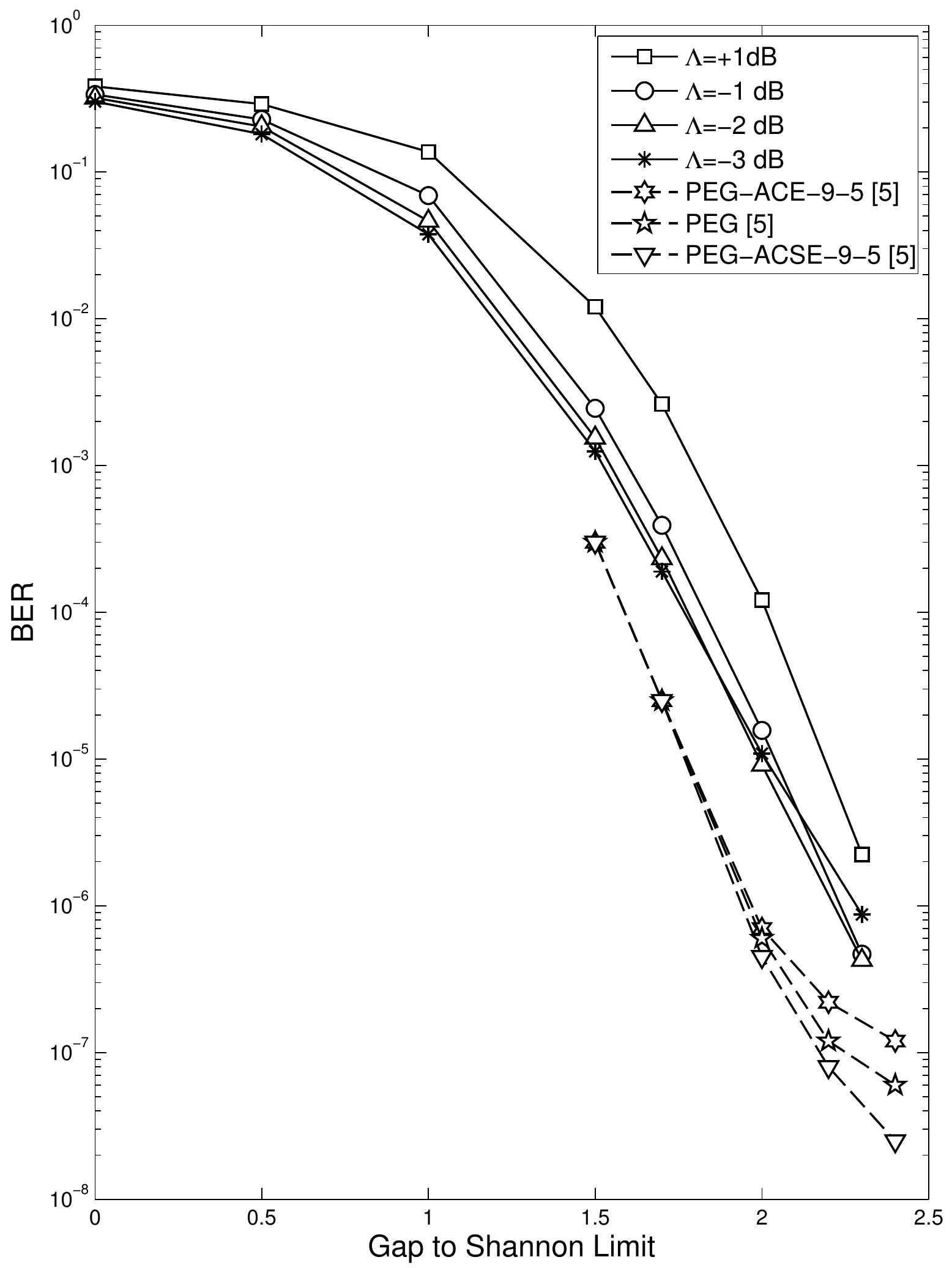}\label{fig:BERq5}}\\
  \caption{BER performance of the proposed system with different values of $q$.}\label{fig:BER}
\end{figure}

Figure \ref{fig:BER} depicts the end-to-end Bit Error Rate (BER) of the system for different values of $q$ and $\Lambda$. Observe that when $\Lambda$ is set beyond a certain threshold, the waterfall region is produced at high SNRs and with no error floor detected in the simulations. The reason being that in this case, early-crossing points in the associated EXIT chart are obtained at low values of $T_{\mathbf{h}}^q(0)$; according to what was stated in Section \ref{sec:EXIT}, any further increase on the value of $T_{\mathbf{h}}^q(1)$ does not yield any iterative processing gain. Consequently, high error floors are obtained for low values of $\Lambda$. Also note that due to the low values of  $T_{\mathbf{h}}^q(1)$ resulting when $q=2$, the system utilizing this value of $\mathbf{h}$ is clearly outperformed by that using $q=3$, $q=4$ and $q=5$.

Also included in these plots are reference curves corresponding to the PEG, PEG-ACSE and PEG-ACE schemes analyzed in \cite[Fig. 6]{Zheng10}. These schemes are based on carefully designed short-length, irregular LDPC codes with low error floors. Despite the good performance of these schemes in terms of error floor versus waterfall tradeoff, it should be emphasized that any change in the specifications of the underlying application would require replacing the entire encoder and decoder, which might be impractical for a variety of scenarios. By contrast, the coding approach here proposed allows for balancing this tradeoff by simply varying $\Lambda$ at a SNR waterfall degradation of less than 0.5 dB (at $\mbox{BER}=10^{\mbox{-}4}$) in all cases. Furthermore, the tradeoff can be easily set based on the EXIT chart associated to the LC code. 

In addition, both the EXIT chart analysis and the simulation results (Fig. \ref{fig:BERq4} and \ref{fig:BERq5}) suggest that for very low BERs (high SNRs) the proposed system will arise lower error floors than the irregular LDPC code proposed in \cite{Zheng10}.

\section*{Acknowledgments}

The authors would like to thank the Spanish Ministry of Science \& Innovation for its support through the \emph{COMONSENS} (CSD200800010) and \emph{COSIMA} (TEC2010-19545-C04-02) projects.

\bibliographystyle{IEEEtran}
\bibliography{./biblio_CommLetter11_short}

\begin{thebibliography}{1}
\providecommand{\url}[1]{#1}
\csname url@samestyle\endcsname
\providecommand{\newblock}{\relax}
\providecommand{\bibinfo}[2]{#2}
\providecommand{\BIBentrySTDinterwordspacing}{\spaceskip=0pt\relax}
\providecommand{\BIBentryALTinterwordstretchfactor}{4}
\providecommand{\BIBentryALTinterwordspacing}{\spaceskip=\fontdimen2\font plus
\BIBentryALTinterwordstretchfactor\fontdimen3\font minus
  \fontdimen4\font\relax}
\providecommand{\BIBforeignlanguage}[2]{{%
\expandafter\ifx\csname l@#1\endcsname\relax
\typeout{** WARNING: IEEEtran.bst: No hyphenation pattern has been}%
\typeout{** loaded for the language `#1'. Using the pattern for}%
\typeout{** the default language instead.}%
\else
\language=\csname l@#1\endcsname
\fi
#2}}
\providecommand{\BIBdecl}{\relax}
\BIBdecl

\bibitem{LDPC}
T.~J. Richardson and R.~Urbanke, ``{The Capacity of Low-Density Parity-Check
  Codes under Message-Passing Decoding},'' \emph{IEEE Trans. Inf. Theory,},
  vol.~47, pp. 599--618, Feb. 2001.

\bibitem{Turbo}
C.~Berrou, A.~Glavieux, and P.~Thitimajshima, ``{Near Shannon Limit
  Error-Correcting Coding and Decoding: Turbo-Codes},'' in \emph{ICC93}, 1993.

\bibitem{LDPC_EF}
T.~Richardson, ``{Error Floors of LDPC codes},'' in \emph{41st Annu. Allerton
  Conf. Commun., Control, Comput.}, October 2003.

\bibitem{Turbo_EF}
R.~Garello, F.~Chiaraluce, P.~Pierleoni, M.~Scaloni, and S.~Benedetto, ``{On
  Error Floor and Free Distance of Turbo Codes},'' in \emph{ICC01}, June 2001.

\bibitem{Zheng10}
X.~Zheng, F.~C.~M. Lau, and C.~K. Tse, ``{Constructing Short-Length Irregular
  LDPC Codes with Low Error Floor},'' \emph{IEEE Trans. on Comm.,}, vol.~58,
  no.~10, pp. 2823--2834, October 2010.

\bibitem{BICM-ID}
S.~Pﬂetschinger and F.~Sanzi, ``{Error Floor Removal for Bit-Interleaved
  Coded Modulation with Iterative Detection},'' \emph{IEEE Trans. on Wireless
  Comm.}, vol.~5, pp. 3174--3181, Nov.

\bibitem{SPA}
F.~R. Kschischang, B.~J. Frey, and H.-A. Loeliger, ``{Factor Graphs and the
  Sum-Product Algorithm},'' \emph{IEEE Trans. Information Theory}, vol.~47,
  no.~2, pp. 498--519, February 2001.

\bibitem{EXIT}
J.~Hagenauer, ``{The EXIT Chart: Introduction to Extrinsic Information Transfer
  in Iterative Processing},'' in \emph{EUSIPCO04}, September 2004.

\end{thebibliography}

\end{document}